\documentclass[twocolumn,pre,aps,showpacs,a4paper,floatfix]{revtex4}

\usepackage{graphicx}
\usepackage{psfrag}
\usepackage{subfigure}
\usepackage{epsfig}
\usepackage{calc}
\usepackage{bm}
\usepackage{longtable}
\graphicspath{{figures/}}

\newcommand{\kappabar}{\bar{\kappa}}

\newcommand{\rhobf}{\bm{\rho}}

\newcommand{\beq}{\begin{equation}}
\newcommand{\eeq}{\end{equation}}
\newcommand{\beqa}{\begin{eqnarray}}
\newcommand{\eeqa}{\end{eqnarray}}
\newcommand{\nn}{\nonumber}

\begin{document}

\title{Budding and vesiculation induced by conical membrane inclusions}
\author{
Thorsten Auth }
\affiliation{Institut f\"ur Festk\"orperforschung and
Institute for Advanced Simulations, Forschungszentrum J\"ulich,
D-52425 J\"ulich, Germany}
\author{
Gerhard Gompper }
\affiliation{Institut f\"ur Festk\"orperforschung and
Institute for Advanced Simulations, Forschungszentrum J\"ulich,
D-52425 J\"ulich, Germany}


\begin{abstract}
Conical inclusions in a lipid bilayer generate an overall spontaneous
curvature of the membrane that depends on concentration and geometry
of the inclusions. Examples are integral and attached membrane proteins,
viruses, and lipid domains. We propose an analytical model to study
budding and vesiculation of the lipid bilayer membrane, which is based on
the membrane bending energy and the translational entropy of the inclusions.
If the inclusions are placed on a membrane with similar curvature radius,
their repulsive membrane-mediated interaction is screened. Therefore, for
high inclusion density the inclusions aggregate, induce bud formation and
finally vesiculation. Already with the bending energy alone our model
allows the prediction of bud radii. However, in case the inclusions induce
a single large vesicle to split into two smaller vesicles, bending energy
alone predicts that the smaller vesicles have different sizes whereas the
translational entropy favors the formation of equal-sized vesicles. Our
results agree well with those of recent computer simulations.
\end{abstract}

\maketitle

\section{Introduction}
Cell membranes contain large amounts of proteins within or attached
to the lipid bilayer \cite{dupuy08}. The distribution of the proteins 
is not necessarily homogeneous, which can have important
functional consequences. For example, proteins with an intrinsic curvature
couple to the bilayer conformation
\cite{sens08,drin07,blood06,auth05,bickel02,ford02,harden94}; on the one
hand, such proteins are preferably found on similarly curved parts of the
membrane \cite{hagerstrand06}, on the other hand, the proteins
deform the membrane locally \cite{kozlov07,antonny06}. Asymmetric, curved
proteins can regulate the polymerization of the three-dimensional
cytoskeleton of the cell \cite{shlomovitz07} and control intracellular
transport via endocytosis \cite{roemer07,mcmahon05}. Virus
endocytosis can occur via the same mechanism \cite{gao05,gozdz07}. The conical
inclusions in our model mimick asymmetric proteins within the bilayer
\cite{ford02}, proteins or polymers attached to the bilayer
\cite{blood06,auth03,tsafrir03}, curved lipid domains
\cite{campelo07,baumgart03,haluska02,gozdz01}, and viruses that bind to the
membrane \cite{gao05}.

The interaction between the inclusions in a lipid bilayer is mediated by
membrane deformations and thermal undulations \cite{bruinsma96,goulian96}, in
addition to surface tension \cite{sens08} and possible direct interactions
that we do not consider in this paper. The deformation-induced, pairwise
interaction of curved inclusions occurs in the absence of thermal membrane
undulations and is usually repulsive \cite{goulian93,goulian93a}; in a planar
membrane it is long-range \cite{weikl98, goulian93, goulian93a}. However,
the interactions can be strongly screened if the average curvature of the
membrane and the protein curvature are similar
\cite{chou01,kim98,dommersnes98}. One obvious example for strongly screened
interactions are inclusions that are placed on a vesicle
with similar curvature radius \cite{dommersnes98}. Screening can also be
achieved by many-body interaction in clusters of inclusions \cite{kim99,kim98}.
At finite temperature, Casimir-like interactions due to membrane undulations
generate attraction \cite{helfrich01,weikl01,golestanian96a,golestanian96b,
goulian93,goulian93a}.

Curvature generation by inclusions and induced budding in lipid bilayer
membranes has been reported in many experimental studies of biological and
biomimetic systems \cite{ford02,antonny06,roemer07,mcmahon05,tsafrir03}.
Computer simulations allow to study the membrane-mediated interaction
between the inclusions in detail without the presence of other, direct
interactions. Recently, bud formation by curved inclusions has
been investigated with computer simulations \cite{reynwar07,
atilgan07}. It was found that the inclusions on the buds have a
higher density than they had in the initially nearly flat membrane
\cite{reynwar07}. This might appear to be a
result of undulation-induced attraction that in consequence leads to
clustering of the inclusions and to budding.

Such systems and processes can be studied theoretically on the basis of
an elastic membrane that is characterized by its bending rigidity, $\kappa$,
and Gaussian saddle splay modulus, $\kappabar$, with curved inclusions
that consist of sections of a sphere with a given opening angle. We
demonstrate that bud formation can already be well understood on the
basis of the membrane deformation alone. We show that the higher
inclusion density on the bud is a result of a screened repulsive interaction.
We further argue that the budding pathway
plays an important role for bud size. This allows us to predict a
range of possible bud radii for a given system, which nicely agrees with
recent simulation results \cite{reynwar07}.

At finite temperature, the inclusions can exist in a fluid and in a
crystalline phase, which depends on the strength of their repulsive
interaction. We construct an approximate free-energy functional that
takes into account for the bending energy as well as the translational
entropy of the inclusions. We calculate a phase diagram for the fission
of a single vesicle of given size and for given number and geometry of
the inclusions. The inclusion entropy plays a decisive role
for the sizes of the smaller vesicles into which a larger
vesicle may split.

\section{Membrane bending energy}
\label{sec2}


\subsection{Membrane shape near inclusions in a lipid bilayer}

The bending energy $\mathcal{E}$ of a lipid bilayer is given by the integral
over the entire membrane area,
\begin{equation}
\mathcal{E} = \int d S \, (2 \kappa H^2 + \kappabar K) \, ,
\end{equation}
where $\kappa$ is the bending rigidity, $\kappabar$
is the saddle-splay modulus, $H=(c_1+c_2)/2$ is the mean curvature,
$K=c_1 c_2$ is the Gaussian curvature, and $c_1$ and $c_2$ are the principal
curvatures at each point of the membrane. The integral over the
Gaussian curvature is determined by the topology of the membrane and by
the geodesic curvature at the boundary. In our case, the geodesic
curvature is given by the geometry of the inclusions, so that in general
this term of the
integral over the membrane shape does not need to be calculated
explicitly. For bud formation, we neglect the constant contribution
of the Gaussian saddle splay modulus.

\begin{figure}[bp]
  \begin{center}
    \leavevmode
    \includegraphics[width=\columnwidth]{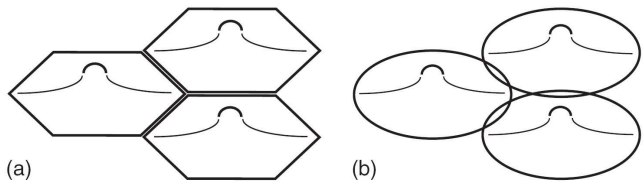}
    \vspace{-3ex}
  \end{center}
    \caption{(a) Membrane deformations induced by curved inclusions
             in a planar membrane. The inclusions have a
             repulsive interaction potential that decreases with the
             distance between the inclusions, $d$, like $V \sim d^{-2}$.
             To minimize the
             bending energy, the inclusions order in an hexagonal
             structure. (b) The hexagons are approximated by overlapping
             circles that have the same projected area.}
    \label{fig3}
\end{figure}

In order to minimize the bending energy, the inclusions preferably order
on a hexagonal lattice (Fig.~\ref{fig3}~{\em (a)}); therefore it is
a natural assumption that the symmetry axis is oriented normal to the local
tangent plane of the vesicle on which the inclusions are placed. To calculate
the deformation energy, we approximate the hexagons with overlapping circles
that have the same projected area (Fig.~\ref{fig3}~{\em (b)}).

If there are no overhangs, the
membrane conformation can be described in Monge parametrization by a
height field, $h(x,y)$, over a planar reference surface. For an almost
planar membrane, the bending energy of the membrane is
\begin{eqnarray}
\mathcal{E} = \frac{1}{2} \kappa \int d A \left[ \Delta h(\rhobf)
\right]^2 \hspace{3ex} (\rhobf=(x,y)) \, ,
\end{eqnarray}
with $\int d A$ the integral over the reference plane. Minimization
of the bending energy gives the biharmonic Euler-Lagrange equation,
\begin{equation}
\Delta^2 h ( \rhobf ) = 0 \, \, \, . \label{eq3}
\end{equation}
In cylindrical coordinates, the general solution of Eq.~(\ref{eq3}) is
\begin{equation}
h(\rho) = \frac{1}{4} \rho^2 ( 2 C_2 - C_3 ) + C_4 + (C_1 +
\frac{1}{2} \rho^2 C_3 ) \ln(\rho) \label{eq7}
\end{equation}
with the four integration constants $C_1$ to $C_4$ \cite{footnote13}.

\begin{figure}[bp]
  \begin{center}
    \leavevmode
    \includegraphics[width=0.80\columnwidth]{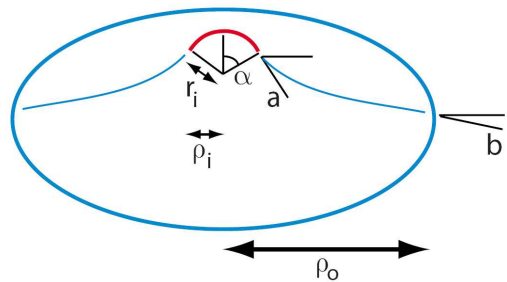}
    \vspace{-3ex}
  \end{center}
    \caption{(Color online)
    Curved inclusion (red) and resulting membrane deformation (blue).
    The inclusion geometry is characterized by the curvature radius, $r_i$, the
    opening angle, $\alpha$, and the projected inclusion radius,
    $\rho_i = r_i \sin(\alpha)$.
    The size of the corresponding membrane patch is $\rho_o$, the slope of the
    membrane at the inclusion is $a = - \tan (\alpha)$, and the slope of the
    membrane at the outer boundary is $b$ ($b=0$ for inclusions on a planar
    membrane).}
    \label{fig4}
\end{figure}

The boundary conditions that are
imposed on the membrane are sketched in Fig.~\ref{fig4}. The radius
of the inner boundary, $\rho_i=r_i \sin(\alpha)$, and the slope of
the membrane at the inner boundary, $h'(\rho_i) \equiv a=-\tan(\alpha)$, are
determined by the inclusion geometry. For $n \approx 4 \pi R^2
\sigma$ inclusions on a vesicle with radius $R$ and surface number
density $\sigma$ of the inclusions, the radius of the
outer boundary is $\rho_o \approx R \sin(\beta)$ with
$\beta=\arccos((n-2)/n)$; the slope of the membrane at the outer
boundary is $h'(\rho_o) \equiv b=-\tan(\beta)$. For inclusions on a planar
membrane, the latter expressions simplify to $\rho_o \approx
1/(\pi\sigma)^{(1/2)}$ and $b=0$. The remaining two boundary
conditions are given by fixing the membrane height at the inner (or
equivalently at the outer) boundary and minimizing the energy with respect
to the height of the inclusion above the vesicle (i.~e.\ the height
difference between both boundaries), which implies $h (\rho_i) = 0$ at the
inclusion and $\partial_\rho \Delta h(\rho) |_{\rho_o} = 0$ at the
outer boundary.

Eq.~(\ref{eq7}) together with the boundary conditions gives the shape
of the deformation,
\begin{equation}
h(\rho) = \frac{(\rho^2-\rho_i^2)(b \rho_o - a \rho_i)+2 \rho_o
\rho_i (a \rho_o-b\rho_i) \ln(\rho/\rho_i)}{2 (\rho_o^2- \rho_i^2)} \, ,
\label{eq10a}
\end{equation}
and the corresponding bending-energy cost, 
\begin{equation}
\mathcal{E} (\rho_o,b) = \frac{\kappa}{2} \int_{\rho_i}^{\rho_o} d
\rho \left[ \Delta_r h(\rho) \right]^2 =  \frac{2 \pi \kappa (b
\rho_o - a \rho_i)^2}{(\rho_o^2-\rho_i^2)} \, \, \, . \label{eq10}
\end{equation}
The energy a function of $\rho_o$ and $b$, which depend on the
inclusion density, while all other quantities are intrinsic properties
of membrane and inclusions. For a single inclusion in an infinite planar
membrane, $b=0$ and $\rho_o \rightarrow \infty$, the bending energy vanishes
and the membrane deformation is catenoid-like,
$h(\rho) = a \rho_i \ln (\rho/\rho_i)$. Note that in a pairwise approximation,
the interaction energy for two inclusions in a planar membrane ($b=0$)
decays like $d^{-2}$ for large distances between the inclusions
(large $\rho_o=d/2$).

\subsection{Optimal, low, and high inclusion density}
\label{sec2b}
\label{sec1c}

\begin{figure}[bp]
  \begin{center}
    \leavevmode
    \includegraphics[width=0.9\columnwidth]{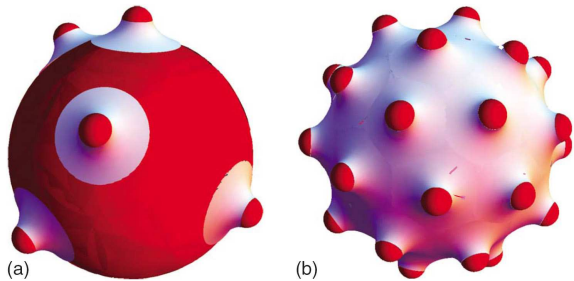}
  \end{center}
    \caption{(Color online)
             (a) Vesicle decorated with curved inclusions. Around each
             inclusion, the membrane can be modeled by segments of the
             catenoid minimal surface (white). The total bending energy
             is $\mathcal{E}=8 \pi \kappa
             (1-S_{\rm cat}/S_{\rm sph})$, where $S_{\rm cat}/S_{\rm sph}$
             is the area fraction of the vesicle that is covered
             with inclusions and catenoidal patches. (b) Vesicle decorated
             with inclusions at optimal density; the bending energy of the
             lipid bilayer membrane vanishes.}
    \label{fig2}
\end{figure}

For inclusions on a vesicle, the membrane shape and the minimal bending
energy (assuming that the inclusions have maximal mutual distances)
can be calculated using Eqs.~(\ref{eq10a}) and (\ref{eq10}). For
$b \rho_o = a \rho_i$, the membrane around the inclusion has almost
catenoid shape \cite{footnote1}; the catenoid is a minimal surface without
bending-energy cost. If the entire vesicle is covered with inclusions and
catenoids such that the bending energy is zero (Fig.~\ref{fig2}~{\em (b)}),
the inclusions have optimal density.

For lower inclusion densities, in a first approximation the catenoid\
shape borders on a spherical shape with the curvature radius of the vesicle.
The bending energy of a vesicle that is decorated with curved inclusions
is reduced by the fraction of the sphere's surface area that is covered
by the inclusions and the catenoid-shaped membrane segments,
see Fig.~\ref{fig2}. Therefore for low inclusion density,
the bending energy of the decorated spherical vesicle is in the range $0
< \mathcal{E} < 8 \pi \kappa$.
In the full solution, which is given by Eq.~(\ref{eq10a}) and will be used
in the remainder of the paper, there is no jump in the mean curvature
from $H=0$ to $H=1/R$ between the catenoid and a sphere as sketched in
Fig.~\ref{fig2}  but rather a smooth transition from zero to finite mean
curvature.

For inclusion densities that are higher than the optimal density, due to the
boundary conditions no solution can be constructed by matching
of catenoids. In this case, the bending energy always
has a finite value that can exceed the bending energy of a bare vesicle.

\begin{figure}[bp]
  \begin{center}
    \leavevmode
    \includegraphics[width=1.0\columnwidth]{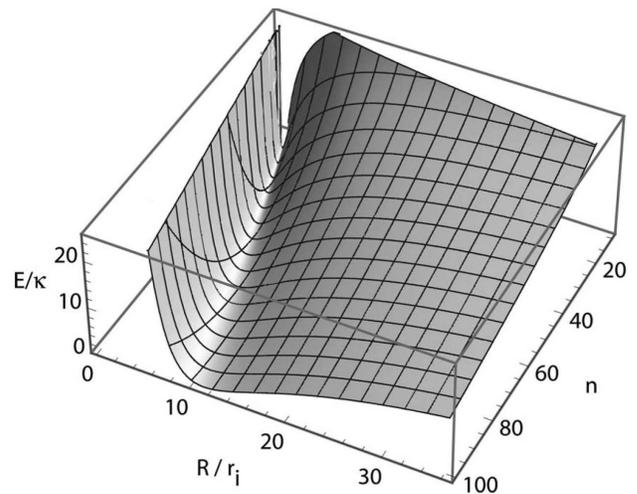}
    \vspace{-3ex}
  \end{center}
    \caption{Normalized bending energy, $E/\kappa$, of a vesicle with
    radius $R$ with $n$
    inclusions ($r_i\approx 5.5 \, \rm nm$, $\alpha=0.64$). There
    is a region of low inclusion density at large $R$ with
    $0 < \mathcal{E} < 8 \pi \kappa$, which is delineated by a
    line of zero-energy ground states from a region of high inclusion density
    at small $R$, where also bending energies $\mathcal{E} > 8 \pi \kappa$
    can be found. (The high energies that are cut off at small $n$ and
    large $R$ mark the breakdown of the small-curvature expansion of the
    bending energy.)}
    \label{fig5}
\end{figure}

The minimal bending energy of a vesicle with inclusions is shown in
Fig.~\ref{fig5} as function of the number of inclusions, $n$, and the vesicle
radius, $R$. We find degenerate zero-energy ground
states that have optimal inclusion density with an approximately linear
dependence $R(n)$, where \cite{footnote2}
\begin{eqnarray}
R \hspace{-1ex} & \approx & \hspace{-1ex} |a| \rho_i n / 4 \approx
1/(\pi \sigma |a| \rho_i) = (\cos
\alpha)/(\pi \sigma \sin^2 \alpha)(1/r_i) \nn \\
& & \hspace{-4ex} {\rm and} \label{eq11} \\
n \hspace{-1ex} & \approx & \hspace{-1ex} 4/(\pi \sigma a^2
\rho_i^2) = (4 \cos^2 \alpha)/(\pi \sigma \sin^4 \alpha)(1/r_i^2) \nn
\end{eqnarray}
The natural spontaneous curvature of the bilayer for given inclusion
density and geometry is $c_0 = 1/R_0 \approx \pi \sigma |a| \rho_i$.

The same value for the inclusion density can be high, optimal, or low,
depending on the radius of the vesicle on which the inclusions are
placed. The smaller the radius of the vesicle, the larger the value
of the optimal density. In a planar membrane, the slopes of two adjacent
catenoid-like deformations cannot be matched for any finite distance
between the inclusions. Therefore the inclusions are always in the
high-density regime in this case.

\subsection{Budding and vesiculation}

\begin{figure}[bp]
  \begin{center}
    \leavevmode
    \includegraphics[width=0.9\columnwidth]{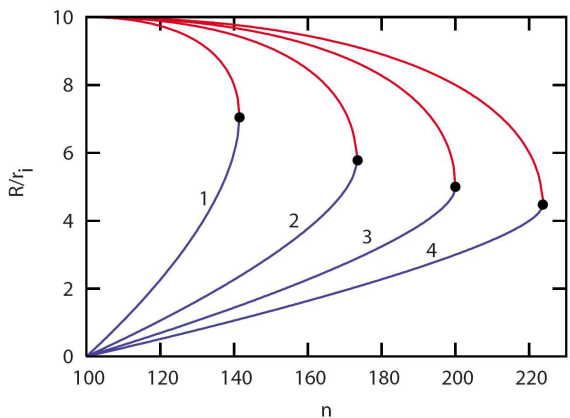}
    \vspace{-3ex}
  \end{center}
    \caption{(Color online) A single vesicle of radius $R=10 \, r_i$ with
    $n$ inclusions splits into one large 'mother'-vesicle and
    several small 'daughter' vesicles, in the figure the cases of 1, 2, 3, and
    4 daughter vesicles are shown. The same parameters as in Fig.~\ref{fig5}
    are used. For 100 inclusions, the initial vesicle
    has vanishing bending energy. For more than 100 inclusions, the sizes
    of mother and daughter vesicles are plotted. The upper branch always
    gives the radius of the mother vesicle, the lower branch is the size of
    the daughter vesicles.
    For a fixed number $n_{\rm v} - 1$ of daughter vesicles (as indicated),
    there is a maximum number of inclusions that allows the formation of
    a state with vanishing bending energy (for which mother and daughter
    vesicles have equal sizes; filled circles).}
    \label{fig5a}
\end{figure}

Bud formation does not occur for a vesicle with low inclusion density
and bending energy,
$0 < \mathcal{E} < 8 \pi \kappa$, because this would lead to an increase of
the total bending energy \cite{footnote3}. However, for high inclusion
density ($n \gtrsim 4 R/(|a| \rho_i)$, see
Eq.~(\ref{eq11})), the system can always reach a state of lower bending
energy if small vesicles bud from the main vesicle.
The set of smaller vesicles into which a large
vesicle with high inclusion density splits up is not uniquely determined
from bending energy alone,
because the states of vanishing bending energy are degenerate. A natural
assumption is that the vesicle will split into one large 'mother' vesicle
and one or more small 'daughter' vesicle(s) of equal size, such that the total
bending energy vanishes and the membrane area is kept constant.
In Fig.~\ref{fig5a}, we show the radii of the mother and daughter vesicles
as function of the number of inclusions.
For a given number of $n_v-1$
daughter vesicles, there is a maximal number of inclusions $n_{\rm max} =
n_v^{1/2} w R$ ($w = 4/(|a| \rho_i)$) that still allows to obtain a zero
energy state, for which mother and daughter vesicles have equal sizes.

If the system can split up into $n_v$ smaller vesicles, it can also split
up into a larger number of small vesicles \cite{footnote4}.
For a vesicle with total number of inclusions $n = n_+ + (n_v - 1) n_-$
and radius $R = (R_+^2 + (n_v - 1) R_-^2)^{1/2}$, bending energy minimization
predicts for the radii and inclusion number on mother ($R_+$, $n_+$) and
daughter ($R_-$, $n_-$) vesicles:
\begin{eqnarray}
n_+ & = & \frac{n + (n_v-1)^{1/2} (n_v w^2 R^2 - n^2)^{1/2}}{n_v} \label{eqvesbr} \\
n_- & = & \frac{n - (n_v-1)^{-1/2} (n_v w^2 R^2 - n^2)^{1/2}}{n_v} \nn \\
R_+ & = & \frac{n + (n_v-1)^{1/2} (n_v w^2 R^2 - n^2)^{1/2}}{n_v w} \nn \\
R_- & = & \frac{n - (n_v-1)^{-1/2} (n_v w^2 R^2 - n^2)^{1/2}}{n_v w}
\, . \nn
\end{eqnarray}
Because of the degeneracy of the states with vanishing bending energy,
thermal fluctuations and the budding pathway play a decisive role to
determine how a large vesicle with high inclusion density splits up into
smaller vesicles. Note that the results of our analytical model so far do
not depend on the value of the bending rigidity of the bilayer.

\subsection{Inclusion clusters}
\label{sec3d}

\begin{figure}[bp]
  \begin{center}
    \leavevmode
    \includegraphics[width=0.9\columnwidth]{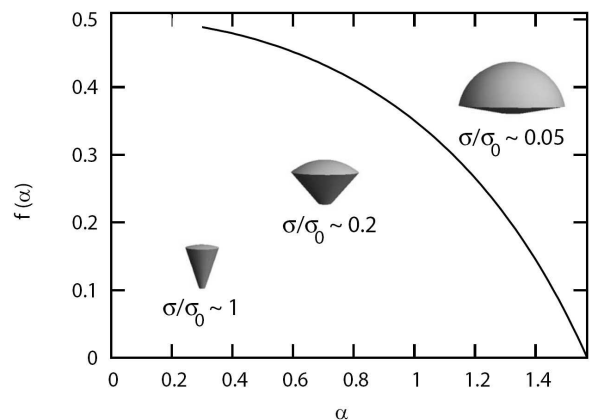}
    \vspace{-3ex}
  \end{center}
    \caption{The coagulation factor, $f(\alpha)=R(\alpha) \pi r_i \sigma_0$,
    describes the dependence of the optimal
    vesicle radius, $R_0$, on the degree of aggregation of the inclusions
    in the membrane. The total inclusion area is kept constant
    and the reference density $\sigma_0$
    corresponds to $\alpha_0=0.1 \, \pi$.}
    \label{fig6}
\end{figure}

A direct attractive interaction between inclusions can induce cluster
formation \cite{sieber07}. In this case, the preferred curvature radius,
$R_0$, not only depends on the number and geometry of the inclusions, but
also on cluster size. For given inclusion curvature radius $r_i$, opening
angle $\alpha_0$, and fixed density $\sigma_0$, inclusion clusters with a
larger opening angle $\alpha$ and reduced density,
$\sigma=(1 - \cos \alpha_0)/(1 - \cos \alpha) \, \sigma_0$, have a stronger
effect on the curvature of the membrane than homogeneously distributed
inclusions. The preferred curvature radius for 
clusters with opening angle $\alpha$ is $R (\alpha) = ((1 - \cos \alpha) \cos
\alpha)/((1- \cos \alpha_0) \sin^2 \alpha)/(\pi r_i \sigma_0) \equiv
f(\alpha)/(\pi r_i \sigma_0)$. We call the normalized curvature radius
$f(\alpha)$ the coagulation factor, because it multiplies 
the curvature radius for
the reference inclusion density and opening angle $\alpha_0$
\cite{footnote6,footnote7}, see Fig.~\ref{fig6}.

For a fixed number $n_0$ of inclusions, the preferred curvature radius
decreases when aggregates are formed,
i.~e.\ the efficiency with which the inclusions influence the membrane
curvature increases. Cluster formation therefore also shifts the high-density
regime to smaller numbers of inclusions for the
same vesicle radius, $n_0 \gtrsim f(\alpha) \, 4 R/r_i$, and may cause a large
vesicle to break up into smaller vesicles.

\section{Thermal fluctuations}

For finite temperature, the translational entropy of the inclusions
contributes to the free energy. We distinguish a crystalline hexagonal
phase and a disordered fluid phase for which we construct free energy
functionals. Phase diagrams have been calculated more rigorously in
Ref.~\cite{netz95} in the limiting case of an almost planar membrane and
for inclusions that are slightly stiffer than the membrane and weakly curved
--- but not in the context of budding.

We neglect the interaction between inclusions by thermal membrane
undulations. For a pair of inclusions and an arbitrary orientation of
the inclusion axis, to lowest order of $\rho_i^2/d^2$ the deformation
energy is
\cite{weikl98,fournier97,goulian93,goulian93a}
\begin{equation}
\mathcal{E}_{\rm def.} = 8 \pi \kappa \alpha^2 \frac{\rho_i^4}{d^4}
\end{equation}
and the undulation-induced interaction energy
\cite{weikl02,golestanian96a,golestanian96b,goulian93}
\begin{equation}
\mathcal{F}_{\rm und.} = - 6 k_B T \frac{\rho_i^4}{d^4} \, .
\end{equation}

The ratio of the membrane deformation-induced repulsion to the
undulation-induced attration in a planar membrane is
$4 \pi \kappa \alpha^2/(3 k_B T)$. The undulation-induced
attraction can be neglected if it is one order of magnitude
smaller than the deformation interaction; for $\kappa=10 \, k_B T$
this is the case for $\alpha \gtrsim 0.5$, for $\kappa = 20 \, k_B T$
already for $\alpha \gtrsim 0.35$.
For low inclusion densities, the undulation-induced interaction
energy decays with the square of the density while the free
energy due to the inclusion entropy is of the order of $k_B T$; in case
of the phase diagrams in Fig.~\ref{fig12}, the undulation free energy
is about $~10^{-4} \, k_B T$.

\subsection{Inclusion effective pair potential and effective hard-disc radius}

For inclusions on a hexagonal lattice, each inclusion corresponds to
three pair interactions and a radius of the deformation patch,
$\rho_o$, that is half the distance between the inclusions. The
effective pair potential, obtained from Eq.~(\ref{eq10}), is thus
\begin{equation}
\begin{array}{c@{\hspace{3ex}}l}
  u (d) = \frac{2 \pi \kappa (b \, d - 2 a \rho_i)^2}{3 (d^2 - 4 \rho_i^2)} &
  \text{if $b \, d \le 2 a \rho_i$} \\
  u (d) \approx 0 & \text{if $b \, d > 2 a \rho_i$} \, .
\end{array}
\label{eq9new}
\end{equation}
For inclusions on a planar membrane ($b=0$) and
for large $d$, i.~e.\ for $d \gg \rho_i$, the repulsive interaction
potential decays with a power law, $u \sim d^{-2}$.

To determine the free energy of this system, we employ the method
developed for suspensions of repulsive colloids \cite{barker67}, i.~e.\
we mimic the interaction potential by hard discs with an effective radius, 
$r_{\rm hd}$. The radius is calculated 
from a comparison of the membrane deformation energy with the thermal energy,
$k_B T$. We use a modified Barker-Henderson method that is appropriate for soft
potentials \cite{barker67,footnote7}, 
\begin{equation}
r_{\rm hd} = \frac{1}{2} \int_{0}^{\rho_u} \left\{ 1 - \exp \left[ -
\beta \mathcal{E} (\rho) \right] \right\} d \rho \, , \label{eq14}
\end{equation}
where the upper integral boundary is determined by $u (2 \rho_u)
= k_B T$ \cite{henderson70, watts69}. The effective hard-disc radii
therefore depend on the geometry of the inclusion, the bending rigidity
of the membrane, and on the radius of the vesicle on which the inclusions
are placed.

\begin{figure}[bp]
  \begin{center}
    \leavevmode
    \includegraphics[width=0.9\columnwidth]{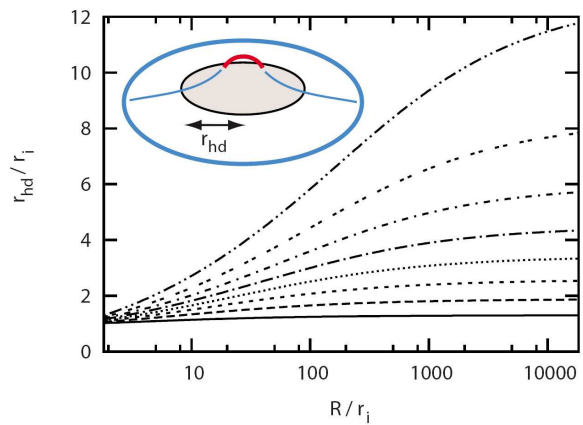}
    \vspace{-3ex}
  \end{center}
    \caption{Effective hard-disc radii for inclusions on a vesicle as
    function of the vesicle's radius, $R$ (see Eq.~(\ref{eq14})). All
    inclusions have the curvature radius, $r_i = 5.5 \, \rm nm$, and
    the opening angles $\alpha=0.4 \pi$, $\alpha=0.35 \pi$, ...,
    $\alpha=0.05 \pi$ (from top to bottom). The projected inclusion
    radii are in the range $0.86 \, \rm nm < \rho_i < 5.2 \, \rm nm$,
    the membrane bending rigidity is $\kappa=12 \, k_B T$.}
    \label{fig7}
\end{figure}

In Fig.~\ref{fig7}, the effective hard-disc radii, $r_{\rm hd}$, are
plotted for inclusions with various opening angles as function of
the vesicle radius, $R$ (an extremely large radius
$R=100 \, \rm \mu m$ of the vesicle is used to describe inclusions
in planar membranes). The hard disc radii increase with
increasing vesicle radius; the increase of $r_{hd}$ with $R$ is the
stronger, the larger the opening angle $\alpha$ of the inclusion is.
For example, the inclusions with $r_i=5.5 \, \rm nm$ and 
$\alpha = 0.4 \, \pi$ can approach each other by diffusion about an order of
magnitude closer on a vesicle with radius $R = 10 \, \rm nm$ than this
is possible on a planar membrane. Consequently, the translational entropy
of the inclusions lowers the free energy on the vesicle compared
with the planar membrane.

\begin{figure}[tp]
  \begin{center}
    \leavevmode
    \includegraphics[width=0.9\columnwidth]{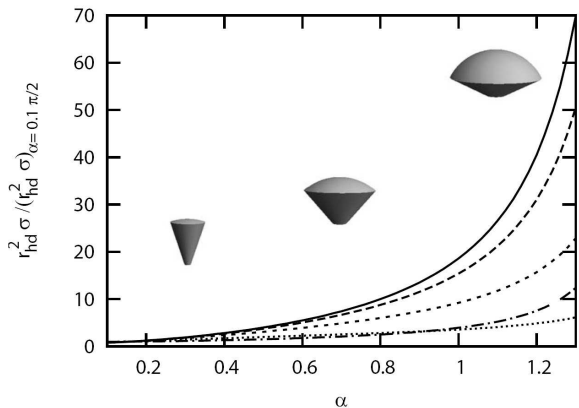}
    \vspace{-5ex}
  \end{center}
    \caption{Area fraction of effective hard discs for inclusions with
    $r_i=5.5 \, \rm nm$ and opening angles $0.1 < \alpha <
    1.3$. For a fixed number of inclusions,
    cluster formation leads to a larger area fraction of the effective
    hard discs and finally to crystallization. The area fraction of the
    effective hard discs with opening angle $\alpha$ is normalized by the
    area fraction of effective hard discs with $\alpha=0.05 \, \pi$: an
    increase of $\alpha$ corresponds to a decrease of $\sigma$ (compare
    Fig.~\ref{fig6}). The inclusions are placed on vesicles with
    $\kappa=12 \, k_B T$ and various radii $R=0.1 \, \rm \mu m$ (dotted),
    $R=1 \, \rm \mu m$ (short-dashed), $R=10 \, \rm \mu m$
    (long-dashed), and $R=100 \, \rm \mu m$ (solid). For
    comparison, the area fraction of effective hard discs is also shown
    for $\kappa = 1 \, k_B T$ and $R=100 \, \rm \mu m$ (dashed-dotted).}
    \label{fig8}
\end{figure}

As discussed in Sec.~\ref{sec3d}, cluster formation of inclusions
increases their effect on the membrane curvature. The effect of cluster
formation on the area fraction of effective hard discs is plotted in
Fig.~\ref{fig8}. For $\kappa=12 \, k_B T$, which is a typical
value for a lipid bilayer, clustering on vesicles with large radii, $R$,
strongly increases the area fraction of the effective hard discs.
To illustrate the strong effect of the bending
rigidity on the effective hard-disc radius, which enters the calculation
of the radius via the exponential function in Eq.~(\ref{eq14}), we 
plot the effective hard disc radii for $\kappa=1 \, k_B T$;
the increase of the area fraction of the hard discs with $\alpha$ is
much smaller than for $\kappa=12 \, k_B T$. Thus the translational entropy
of the inclusions plays a much more important role for smaller
$\kappa$ \cite{footnote8}.

\subsection{Inclusion entropy and free energy of hard discs}

The free energy of a fluid of hard discs can be very well described
by the Carnahan-Starling
free energy \cite{carnahan69,maeso93}. It is the sum of the ideal-gas free
energy, $\mathcal{F}_{\rm id}/n \approx k_B T \log(\sigma \Lambda^2)$,
where $\Lambda$ is the thermal wavelength (see e.~g.\ Ref.~\cite{goegelein08}),
and the excess free energy $\mathcal{F}_{\rm CS}$ \cite{hansenbook}. The
latter is calculated from the
Carnahan-Starling equation of state \cite{maeso93},
\begin{equation}
\frac{p}{\sigma \, k_B T} = \frac{1}{(1-y)^2} \, \, \, ,
\end{equation}
with the hard-disc area fraction, $y=\sigma \pi r_{\rm hd}^2$.
Integration of the thermodynamic relation $p=-(\partial
F/\partial V)_{T,N}$ finally gives the excess free energy,
\begin{equation}
\frac{1}{n} \frac{\mathcal{F}_{\rm CS}}{k_B T}  = \int_0^{y} \left(
\frac{p}{k_B T \, \sigma} - 1 \right) \frac{d\tilde{y}}{\tilde{y}} =
\frac{y}{1-y} - \ln (1-y) \, \, \, .
\end{equation}
The Carnahan-Starling excess free energy diverges at the crystallization
transition of the effective hard discs.

Because in the fluid as well as in the crystalline phase of the inclusions
the squared thermal wavelength enters through the same constant and additive
term, we consistently replace it in both cases --- without loss of
information --- by the projected area of the inclusion, $\pi \rho_i^2$,
such that
$\tilde{\mathcal{F}}_{\rm id}/n \approx k_B T \log(\sigma \pi \rho_i^2)$.

\begin{figure}[hp]
  \begin{center}
    \leavevmode
    \includegraphics[width=0.9\columnwidth]{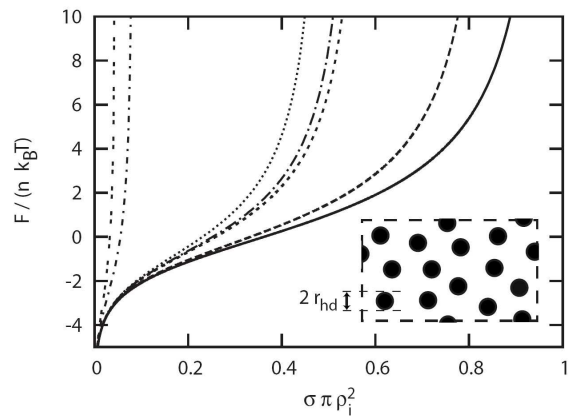}
    \vspace{-3ex}
  \end{center}
    \caption{Carnahan-Starling free energy of hard discs as function
    of the inclusion area fraction. The system is depicted 
    in the inset with periodic boundary conditions. We plot the free
    energy for discs with the geometrical projected radii of the
    inclusions, $r_{\rm hd}=\rho_i = r_i \sin \alpha$ (solid line), as well
    as the free energies for effective hard discs for inclusions with
    curvature radius $r_i = 5.5 \, \rm nm$ and various opening angles
    on vesicles with $\kappa=12 \, k_B T$:
    $\alpha = 0.16$ and $R = 20 \, \rm nm$ (long dashed),
    $\alpha = 0.16$ and $R = 100 \, \rm \mu m$ (short dashed),
    $\alpha = 0.64$ and $R = 20 \, \rm nm$ (long-dashed dotted),
    $\alpha = 0.64$ and $R = 100 \, \rm \mu m$ (short-dashed dotted),
    $\alpha = 0.82$ and $R = 20 \, \rm nm$ (dotted), and
    $\alpha = 0.82$ and $R = 100 \, \rm \mu m$ (double dashed).}
    \label{fig9}
\end{figure}

Usually the translational entropy favors a homogeneous distribution of
particles. However, because the effective hard-disc radius depends on
the membrane curvature, on a deformable membrane, a homogeneous distribution
of inclusion does not need to be the most favourable state. Instead, the
inclusion density on the bud can be higher than on the mother vesicle because
of the screened repulsive interactions.
Fig.~\ref{fig9} shows the free energies of a fluid of effective hard
discs with curvature radius $r_i = 5.5 \, \rm nm$ for various opening
angles in a lipid bilayer with $\kappa=12 \, k_B T$.
For nearly identical curvature radii of vesicle and inclusions, the
effective hard-disc radius almost coincides with the geometric hard-disc
radius of the inclusion.

\subsection{Free energy per inclusion in fluid and crystalline phases}
\label{sec4}

\begin{figure}[bp]
  \begin{center}
    \leavevmode
    \includegraphics[width=0.93\columnwidth]{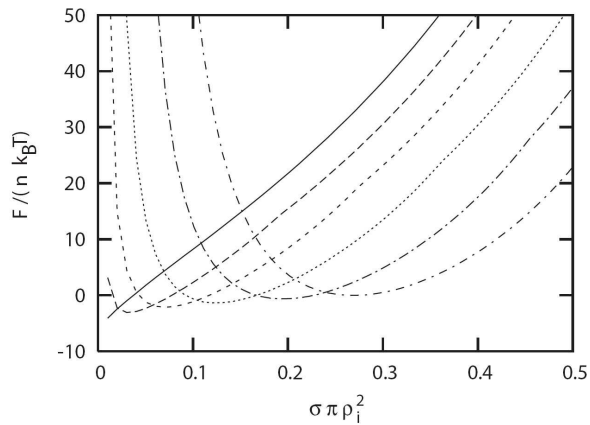}
    \vspace{-3ex}
  \end{center}
    \caption{Free energies as function
    of the inclusion area fraction, $\sigma \pi \rho_i^2$, for inclusions
    with $r_i=5.5 \, \rm nm$, $\alpha = 0.82$ in a membrane with
    $\kappa= 12 \, k_B T$ for several vesicle radii:
    $R = 1 \, \rm \mu m$ (solid),
    $R = 100 \, \rm nm$ (long-dashed), $R = 50 \, \rm nm$ (short-dashed),
    $R = 30 \, \rm nm$ (dotted), $R = 20 \, \rm nm$ (long dashed-dotted),
    $R = 15 \, \rm nm$ (short dashed-dotted). For low densities, the
    inclusions are in a fluid phase and the free energy is given by the
    membrane bending energy plus the Carnahan-Starling excess free energy
    for the effective hard discs. For high densities, the inclusions are in
    a crystalline phase and the free energy is given by the membrane
    bending energy and the free energy of a harmonic crystal.}
    \label{fig10}
\end{figure}

We construct the free energy per inclusion in the crystalline phase
from the membrane
bending energy and the fluctuation free energy of a harmonic crystal,
and in the fluid phase from the sum of the membrane bending energy and
the translational entropy of the inclusions \cite{footnote12}.

For the harmonic crystal, the spring constant $k_{\rm sp}$ is obtained
for a hexagonal lattice with the interaction potential in Eq.~\ref{eq9new},
\begin{equation}
k_{\rm sp} = \frac{16 \pi \kappa \rho_i}{(d^2-4 \rho_i^2)^3} (3 d^2 +4 \rho_i^2) \rho_i (a^2+b^2)-(d^2+12 \rho_i^2) a b d \, . 
\end{equation}
The free energy contribution of the positional fluctuations of the inclusions
is therefore
\begin{equation}
\mathcal{F}_{\rm HC} = k_B T \ln \left[\frac{k_{\rm sp} \Lambda^2}{2 \pi}\right]
\end{equation}
or --- after replacement of the thermal wavelength by the inclusion size ---
\begin{equation}
\tilde{\mathcal{F}}_{\rm HC} = k_B T \ln \left[\frac{k_{\rm sp} \rho_i^2}{2}\right] \, .
\end{equation}
Whenever we use the free energy for the crystalline phase in this paper, the
Lindemann parameter remains below the critical value for melting of
the inclusion crystal
\cite{eisenmann04}.

The transition between the fluid and the crystalline
phase already occurs below the crystallization transition of the effective
hard discs, $\sigma \pi r_{\rm hd}^2 \approx 0.7$. In Fig.~\ref{fig10},
the free energy per inclusion is plotted for several vesicle radii.
Entropy reduces the optimal bud radius for a given inclusion density
compared with Eq.~(\ref{eq11}). However, the bending energy alone still
provides a good estimate for the optimal bud radius, because of the strong
increase of the free energy per inclusion for low inclusion densities
(see Fig.~\ref{fig10}).

\subsection{Vesiculation diagrams}

\begin{figure}[tp]
  \begin{center}
    \leavevmode
    \includegraphics[width=0.75\columnwidth]{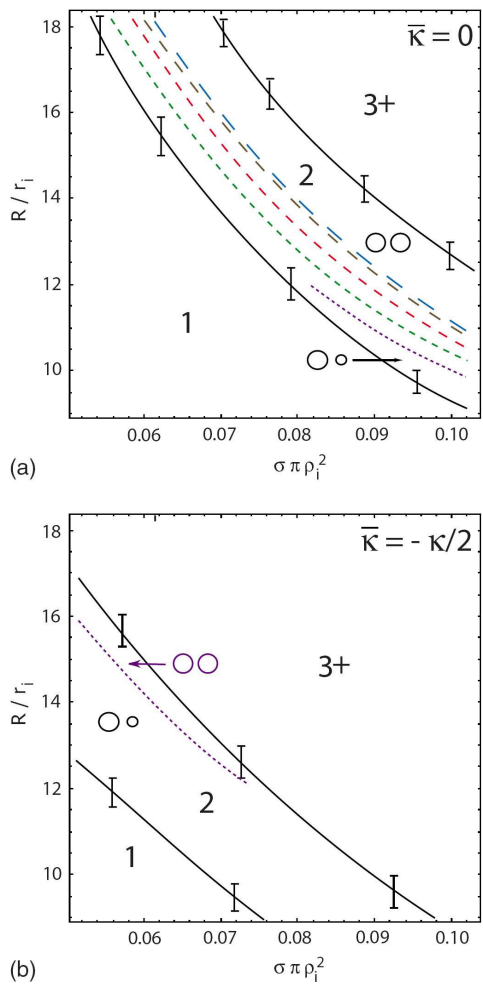}
    \vspace{-3ex}
  \end{center}
    \caption{(Color online)
    Vesiculation phase diagram for an inclusion density $\sigma$
    of inclusions with $r_i=5.5 \, \rm nm$ and $\alpha = 0.64$ on (initially)
    a single vesicle of radius $R$. For small $\sigma$ or $R$, the
    energetically favorable state is the single vesicle; fission, first
    into two, and finally into $3$ or more vesicles is expected to occur
    when $\sigma$ and/or $R$ is increased (more than 3 vesicles
    are not resolved by the calculation). {\em (a) $\kappabar=0$:} 
    In the two-vesicle regime, a region where the two vesicles have equal
    sizes grows with decreasing $\kappa$ and bounds the three-vesicle regime.
    The lines depict the border between equally and differently sized vesicles
    for $\kappa=200 \, k_B T$, $\kappa=100 \, k_B T$, $\kappa=50 \, k_B T$,
    $\kappa=30 \, k_B T$, $\kappa=10 \, k_B T$.
    (the vesicle sizes are not resolved in the $3+$ region).
    {\em (b) $\kappabar=-\kappa/2$:} Fission occurs already for smaller
    values of $\sigma$ and $R$. In the $2$-vesicle region the two
    vesicles have different sizes for $\kappa=200 \, k_B T$,
    $\kappa=100 \, k_B T$, $\kappa=50 \, k_B T$, and $\kappa=30 \, k_B T$.
    For $\kappa = 10 \, k_B T$, in a small region both vesicles have
    equal sizes.}
    \label{fig12}
\end{figure}

From the total free energy in Sec.~\ref{sec4}, we calculate vesiculation
diagrams starting with a single vesicle of radius $R$ and a given number of
inclusions for several values of $\kappa$. Because the topology changes
when buds detach, the value of $\kappabar$ plays an important role for
vesiculation.
Fig.~\ref{fig12} shows vesiculation diagrams for $\kappabar=0$
and for $\kappabar=-\kappa/2$ (the ratio $\kappabar/\kappa$ is still under
debate, see Ref.~\cite{deserno09}). We calculate whether the single
vesicle is the most stable
state or if fission in two or more smaller vesicles is favorable.
With increasing initial vesicle radius and inclusion density, fission
becomes more likely to occur --- first into two, at even larger $R$ or
$\sigma$ into three or more vesicles.

For $\bar{\kappa} = 0$, fission in the bending-energy dominated regime
produces two smaller vesicles that in general have different sizes,
see Eq.~(\ref{eqvesbr}). If entropy is important, the two
vesicles may have equal size. In Fig.~\ref{fig12}~(a), it is shown
that a regime of equally-sized vesicles develops, bordering the
three-vesicle regime and increasing in size with decreasing $\kappa$.
Within the error bars of our calculation, we find that the boundaries
between one, two, and three vesicles are independent of the
value of the bending rigidity for $10 \, k_B T \le \kappa \le 200 k_B T$.

For $\kappabar = - \kappa/2$, vesiculation takes place
for smaller initial vesicle radii and inclusion densities than for
$\kappabar = 0$, because each additional vesicle lowers the free
energy by $4 \pi \kappabar$. In the entire two vesicle regime, both vesicles 
have different sizes for $\kappa=200 \, k_B T$, $\kappa=100 \, k_B T$,
$\kappa=50 \, k_B T$, and $\kappa=30 \, k_B T$, while for
$\kappa = 10 \, k_B T$ a small region of equally sized vesicles is found,
see Fig.~\ref{fig12}~(b).

Note that for bud formation, which has to occur before vesiculation,
the value of $\kappabar$ is irrelevant and the phase diagrams for
$\kappabar=0$ apply (assuming that the buds are connected by
catenoidal necks with vanishing bending energy, and that the membrane
area needed to form the neck is negligible). While the bud is
being formed and has not yet detached, the integral over the Gaussian
curvature and therefore the contribution of the saddle splay modulus to the
deformation energy stays constant.
However, a negative saddle splay modulus facilitates the neck
between two vesicles to break. In this case, the budded
state can act as energy barrier for vesiculation that prevents
fission, separating a high-energy single-vesicle state and a low-energy
state of several smaller vesicles.

\section{Budding pathway}
\label{sec5}

To shed more light on the role of the budding pathway, we compare the
typical diffusion time of the inclusions with the relaxation time of
the membrane conformation on the same length scale. If the
diffusion of inclusions is fast compared with the relaxation time of
the membrane, the initial membrane shape is decisive; for
a fast membrane relaxation, the initial distribution of inclusions
mainly determines the budding process.

The diffusion time is
$t_d= \lambda^2/(4 D)$ where $\lambda$ is a characteristic
length scale that the inclusion has diffused and $D$ is the diffusion
coefficient of the inclusion. A typical value is $D= 1 \, \rm \mu
m^2 s^{-1}$ for the diffusion of lipids and the diffusion coefficient
for proteins in cell membranes can be up to two orders of
magnitude smaller \cite{henis93}. The relaxation time of the
membrane is $t_r = \eta \lambda^3/(2 \pi^3 \kappa)$ \cite{gov03},
where $\eta=1 \, \rm mPa \, \approx 2.4 \,
10^{-10} \, k_B T \, \rm s \, nm^{-3}$ is the viscosity of the
surrounding water and $\lambda$ is the wavelength of the membrane
undulations. From the cubic versus the quadratic dependence on
$\lambda$, we find that for small $\lambda$, $t_r < t_d$, while for
large $\lambda$, $t_r>t_d$.

For $\kappa= 10 \, k_B T$ and $D=1 \, \rm \mu m^2 s^{-1}$ (which is
an upper bound for the diffusion coefficient of proteins), we find
that $t_r=t_d$ for $\lambda \approx 0.6 \, \rm mm$. This length is
much larger than $10 \, \rm \mu m$, which is the size of cells
\cite{albertsbook} or giant unilamellar vesicles \cite{pecreaux04},
thus the initial aggregation of inclusions is diffusion-limited. We
therefore assume that inhomogeneities in the protein distribution on
the membrane will immediately lead to a membrane deformation that
minimizes the bending energy. The larger the initial inclusion density,
the larger the spontaneous curvature of the membrane (compare
Eq.~(\ref{eq11})) and the smaller the size of the buds that are formed.

Bud formation is initialized in some regions of the membrane that have
a noticeably higher inclusion density than others. The relative protein
density fluctuations decrease with the size of a membrane patch which is
considered. If we assume a random distribution of inclusions,
for a patch with $N$ inclusions the relative fluctuations of the inclusion
number are of the order of $N^{-1/2}$. Thus for small patches, the
inhomogeneities are strongest and budding will preferably occur on the
smallest possible lengthscale. A small average membrane curvature with
appropriate sign further attracts proteins to those regions where the
bending energy per inclusion is already reduced. However, this clustering
of inclusions during the budding process is hindered by a ring with
opposite membrane curvature that forms the neck of a growing bud. This
ring acts as energetic barrier that prevents further inclusions to enter
a patch of the membrane where budding has already started \cite{footnote10}.
Because of the neck formation and the diffusion-limited budding
process, the bud size is roughly determined by the initial
inclusion density on the membrane.

\section{Comparison with simulation results}

\begin{figure}[tp]
  \begin{center}
    \leavevmode
    \includegraphics[width=0.9\columnwidth]{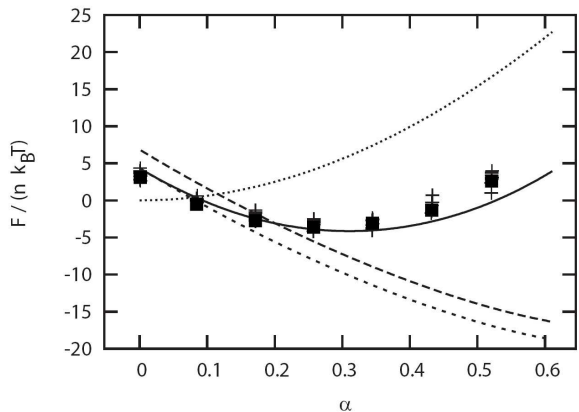}
    \vspace{-3ex}
  \end{center}
    \caption{Energies per inclusion needed to place $10$ inclusions
    of size $\rho_i=2.5 \, \rm nm$ on a vesicle with $R = 15 \, \rm nm$,
    $\kappa = 20 \, k_B T$, and $\kappabar = - 20 k_B T$, as function
    of $\alpha$. Lines show bending energy (long-dashed), bending energy
    and inclusion
    entropy (short-dashed), saddle-splay energy (dotted), and total energy
    (solid). For comparison, we also plot the simulation data of
    Ref.~\cite{atilgan07} (symbols indicate different calculation methods 
    \cite{atilgan07}), shifted by $\Delta F = -10.5 \, k_B T$ (see main text).
    The deviation of the simulations and our theory
    for $\alpha \gtrsim 0.4$ might be due to the surface tension used
    in the simulation, which is not included in our model.}
    \label{fig11}
\end{figure}

Budding due to membrane inclusions has been studied recently with computer
simulations \cite{reynwar07,atilgan07}. In Ref.~\cite{atilgan07}, entire
vesicles with inclusions are simulated where the membrane is modelled as a
dynamically triangulated surface. In Ref.~\cite{reynwar07}, coarse-grained
model lipids are used to study budding for planar bilayer patches.
 
In Fig.~\ref{fig11}, the different contributions to the free energy
per inclusion needed to graft $10$ inclusions with given projected
radius $\rho_i = 2.5 \, \rm nm$ on a vesicle with radius $R=15 \,
\rm nm$ are plotted as function of the opening angle, $\alpha$. The
inclusions are in the fluid phase.
For comparison, we also plot the simulation data of Ref.~\cite{atilgan07},
shifted by $\Delta F = -10.5 \, k_B T$ because our model does not account
for thermal undulations of the membrane conformation. This energy difference
is extracted from the simulation data for $T= 300 \, \rm K$ and for
$T = 3 \, \rm K$, see Fig. 6A in Ref.~\cite{atilgan07}. However, the very
good match is somewhat fortuituous because we replace the thermal wavelength
in the ideal gas free energy by the inclusion size, such that it is similar
to the free energy obtained on a triangulated vesicle with a bond length
that approximately equals to the inclusion diameter, compare \cite{smith05}.

We consider curvature radii of the inclusions that are both smaller and larger
than the curvature radius of the vesicle. For $\alpha \rightarrow 0$ and $r_i
\rightarrow \infty$, such that $\rho_i = 2.5 \, \rm nm$, bending
energy is needed to insert the flat inclusion into the curved
vesicle. This bending energy cost decreases with increased
opening angle $\alpha$ and is zero for $\alpha \approx 0.17$ where the
curvature radius of the inclusion equals the curvature radius of the
vesicle. If $\alpha$ is further increased, the bending energy per
inclusion continues to decrease as more and more of the vesicle area
consists of catenoidal patches around the inclusions. For $\alpha \approx
1.18$, i.~e.\ for even larger opening angles than plotted in the figure,
the inclusions have optimal density and the bending energy
gained by grafting all $10$ inclusions to the vesicle is $8 \pi
\kappa$.

In addition to the bending energy, there is an energy
cost $\mathcal{E}_{\bar{\kappa}}$ for grafting that arises due to
the saddle-splay modulus, which has been
chosen to be $\kappabar = - \kappa = - 20 \, k_B T$ for consistency
with Ref.~\cite{atilgan07}. The energy cost per inclusion only
depends on the geodesic curvature of the membrane at the inclusion,
i.~e.\ on the opening angle $\alpha$, which implies
$\mathcal{E}_{\kappabar} = - 2 \pi \kappabar (1 - \cos \alpha)$. 
Therefore it is independent of vesicle
radius and inclusion density and is only important to calculate the
chemical potential for the inclusions on the surface; the budding
transition at given inclusion density in the membrane
is independent of $\bar{\kappa}$.

In the simulations presented in Ref.~\cite{reynwar07}, budding is induced
by inclusions with $r_i = 5.5 \, \rm nm$ that are initially
placed in a regular array on a planar membrane with $\kappa = 12 \, k_B T$.
Under the assumption that the initial inclusion density,
$\sigma = 2 \, 10^{-3} \, \rm nm^{-2}$, determines the bud size
(see Sec.~\ref{sec5}), we can
roughly predict the bud radius from Eq.~(\ref{eq11}).
Possible bud radii are estimated by comparing
the free energies for different vesicle radii in Fig.~\ref{fig10} at
the initial inclusion density.

The parameters for which
the free energies are plotted in Fig.~\ref{fig10} are chosen to
match the bending rigidity and the inclusion geometry of the
'large inclusions' in Ref.~\cite{reynwar07} with $\alpha = 0.26 \, \pi$
\cite{footnote11}. For
an initial inclusion density in the planar membrane, $\sigma=0.002 \,
\rm nm^{-2} \approx 0.15 \, \sigma \pi \rho_i^2$, which is estimated by visual
inspection from the simulation snapshots, we find that the inclusions
are in the crystalline phase on the
vesicle with $R=1 \, \rm \mu m$ (i.~e.\ in a planar membrane).
The free energy per inclusion is about $10 \, k_B T$. Significantly
smaller free
energies per inclusion can be found for vesicle radii
$22 \, {\rm nm} \lesssim R \lesssim 100 \, \rm nm$;
the optimal vesicle radii are $30 \, {\rm nm} \lesssim
R \lesssim 60 \, \rm nm$ with free energies per inclusion of
about $-1 \, k_B T$. These radii agree well with the observed
bud radius of $R=30 \, \rm nm$ that formes in the simulations as final
state via an initially slightly larger bud. The optimal bud radius from
Eq.~(\ref{eq11}) is approximately $37 \, \rm nm$.

Similarly, for the 'very large inclusions' in Ref.~\cite{reynwar07}
($\alpha = 0.39 \, \pi$)
we predict bud radii, $15 \, {\rm nm} \lesssim R \lesssim 20 \, \rm
nm$, as observed in the simulations; based on bending energy only we
find from Eq.~(\ref{eq11}) $R\approx 11 \, \rm nm$. For the
'small inclusions' in Ref.~\cite{reynwar07} ($\alpha = 0.20 \, \pi$)
that are already in the planar
membrane almost in the fluid phase, our model predicts for the $36$
inclusions studied in the simulations a maximal gain for the free
energy per inclusion of $\approx 1.5 \, k_B T$ for $R = 38 \, \rm nm$
and a strong decrease of this energy gain for smaller vesicle radii.
Already for
$R=34 \, \rm nm$ and $29$ inclusions, the energy per inclusion on the
bud and in the plane are approximately equal.
Larger energy gains are possible for larger bud radii, for which
many more than the simulated 36 inclusions and a larger area of the bilayer
patch are needed. From these considerations, it is not surprising that no
budding is observed for the 'small inclusions' in the simulations of
Ref.~\cite{reynwar07}.

\section{Conclusions}

We have calculated the membrane-mediated interaction of conical inclusions
in a lipid bilayer and the inclusion entropy, which allow the prediction of
budding transitions
and vesiculation. Our model is based on the membrane bending energy;
with this contribution alone, the spontaneous curvature of a bilayer
with inclusions as well as budding can be predicted for many 
systems, ranging from protein inclusions to viral budding.
Although the interaction between the inclusions by membrane deformation
is repulsive, the screening of the repulsive interaction due to the
average membrane curvature allows higher inclusion densities
on a bud than in the initial vesicle. Translational entropy of the inclusions
favors the formation of equally-sized daughter vesicles and lifts the
degeneracy that is found for states with vanishing bending energy.

From our calculations, the following picture of the effect of the bilayer
deformation by curved inclusions emerges. For low inclusion density,
the membrane around each inclusion assumes a catenoid
shape of vanishing curvature energy. At optimal inclusion density,
the catenoids are closely packed and the bending energy for the entire
vesicle vanishes. For high inclusion density, the boundary conditions for
the membrane deformation around the inclusion do not allow the formation
of catenoidal patches, and the inclusions always feel the membrane-mediated
repulsive interaction with neighboring inclusions. In this regime, bud
formation can occur.

If we assume a specific biological mechanism that leads to formation of
clusters with well-defined and limited size, we find that such a mechanism
can induce
bud formation and vesiculation without the need to insert additional conical
proteins into the cell membrane: cluster formation reduces the preferred
curvature radius of the membrane. We quantify the effect of aggregation
by the coagulation factor, which describes how the preferred curvature
radius for a fixed amount of inclusions changes with the cluster size.

In general, our analytical model is applicable for a wide range of
length scales and inclusion geometries. Computer simulations are usually
designed only for a specific length scale, e.~g.\ a length scale comparable
to the length scale of lipids in Ref.~\cite{reynwar07} or the lengthscale
of entire vesicles in Ref.~\cite{atilgan07}. The good agreement with the
simulation results of Refs.~\cite{reynwar07,atilgan07} suggests that the
approximations used in our calculations are justified.

We argue that the undulation-induced attraction can be
neglected compared with the deformation-induced repulsion
and the translational entropy of the inclusions for
$\kappa \alpha^2 \gtrsim 15 \, k_B T/(2 \pi)$, i.~e.
$\alpha \gtrsim 0.35$ for $\kappa=20 \, k_B T$. For example,
the BAR protein induces a local membrane curvature with an
opening angle $\alpha \approx 0.4$ \cite{blood06}. Clathrin
can induce a variety of opening angles \cite{kohyama03,heuser80}.

The number of the inclusions per bud is determined by the budding process.
Around a growing bud, a neck forms that presents
an energetic barrier for the diffusion of inclusions. Because the deformation
of the lipid membrane typically occurs much faster than the diffusion of
the inclusions within the membrane, the number of inclusions per
bud is well determined by the initial inclusion density in the
membrane. From this, we can estimate a range of possible bud radii,
which agrees well with the simulations in Ref.~\cite{reynwar07}.

It would be interesting to test
the validity of our model in the limits of (a) very
small inclusions, such as lipids with large headgroups, when the
description of the lipid membrane by curvature
elastic constants may not be appropriate and (b)
very floppy membranes, when neglecting the thermal membrane undulations
may not be justified.

\acknowledgments

We acknowledge helpful discussions with G.~N\"agele, R.~G.~Winkler,
J.~L.~McWhirter, K.~Mecke, R.~Golestanian, M. Deserno, and M. Oettel.


\begin{thebibliography}{66}
\bibitem[{foo({\natexlab{a}})}]{footnote13}
\bibinfo{note}{In general, the boundary conditions are given by
$h ( \rho_b ) = h_0 ( \rho_b )$ and $h'(\rho_b) = h_{1} ( \rho_b )$
for the inner and the outer boundaries each or alternatively ---
if the values of $h$ and $h'$ are not known at the boundary --- by the
natural boundary conditions
$\partial_\rho \Delta h( \rho) |_{\rho = \rho_b} = 0$ and
$\Delta h (\rho) |_{\rho = \rho_b} = 0$, respectively \cite{dan94}.}
\bibitem[{foo({\natexlab{b}})}]{footnote1}
\bibinfo{note}{Part of a small sphere (the inclusion) is matched to the narrow
  part of the catenoid such that the slope is continuous, which again is
  matched to a larger sphere (the vesicle), see Fig.~\ref{fig2}.}
\bibitem[{foo({\natexlab{c}})}]{footnote2}
\bibinfo{note}{The equations are applicable for large $n$, but also in the
  exact relations the inclusion geometry enters only as a single variable, $|a|
  \rho_i$.}
\bibitem[{foo({\natexlab{d}})}]{footnote3}
\bibinfo{note}{Budding in this regime of inclusion concentration might be
  possible for a negative value of $\kappabar$.}
\bibitem[{foo({\natexlab{e}})}]{footnote4}
\bibinfo{note}{Eq.~(\ref{eq11}) is valid for large $n$; for very small vesicles
  the approximation breaks down and in addition discretization effects become
  important.}
\bibitem[{foo({\natexlab{f}})}]{footnote6}
\bibinfo{note}{We neglect the contribution of the geodesic curvature, the
  influence of which is characterized by $\bar{\kappa}$.}
\bibitem[{foo({\natexlab{g}})}]{footnote7}
\bibinfo{note}{In the original method that applies to steep potentials,
  $\mathcal{E} (\rho_u)= 0$ is used.}
\bibitem[{foo({\natexlab{h}})}]{footnote8}
\bibinfo{note}{It is not clear whether for $\kappa \simeq 1 \, k_B T$ the
  thermal undulations of the membrane can still be neglected, i.~e.\ whether
  our model applies.}
\bibitem[{foo({\natexlab{i}})}]{footnote12}
\bibinfo{note}{The cell approximation \cite{doege04} for the free energy of the
  crystalline phase is not useful, because the effective hard discs overlap
  already at densities that are slightly higher than the crystallization
  transition of the inclusions and below densities that are observed in
  simulations.}
\bibitem[{foo({\natexlab{j}})}]{footnote10}
\bibinfo{note}{By this mechanism, the final number of inclusions per bud is
  already determined in an early stage of bud formation. The area of lipid
  bilayer per bud can adjust during the budding process and is regulated by the
  number of inclusions on the bud.}
\bibitem[{foo({\natexlab{k}})}]{footnote11}
\bibinfo{note}{Many numerical values of the parameters in the figures have been
  chosen such that they correspond to those in Ref.~\cite{reynwar07}.}
\end{thebibliography}
\end{document}